\documentclass[twocolumn]{aastex6}
\usepackage{amsmath,amstext,bm}
\usepackage{amssymb}
\usepackage{epsfig}
\usepackage{color}
\usepackage{hyperref}
\usepackage{float}

\begin{document}

\title{A two-zone blazar radiation model for ``orphan'' neutrino flares}

\author{Rui Xue\altaffilmark{1, 2, 3}, Ruo-Yu Liu\altaffilmark{1, 2}$^{,*}$, Ze-Rui Wang\altaffilmark{1, 2},  Nan Ding\altaffilmark{4} and Xiang-Yu Wang\altaffilmark{1, 2}}
\altaffiltext{1}{School of Astronomy and Space Science, Nanjing University, Nanjing 0110093, China; *Email: \textcolor{blue}{ryliu@nju.edu.cn}}
\altaffiltext{2}{Key laboratory of Modern Astronomy and Astrophysics (Nanjing University), Ministry of Education, Nanjing 210023, People's Republic of China}
\altaffiltext{3}{Department of Physics, Zhejiang Normal University, Jinhua 321004, China}
\altaffiltext{4}{School of Physical Science and Technology, Kunming University, Kunming 650000, China}

\begin{abstract} 
In this work, we investigate the 2014-2015 neutrino flare associated with the blazar TXS~0506+056 and a recently discovered muon neutrino event IceCube-200107A in spatial coincidence with the blazar 4FGL~J0955.1+3551, under the framework of a two-zone radiation model of blazars where an inner/outer blob close to/far from the supermassive black hole are invoked. An interesting feature that the two sources share in common is that no evidence of GeV gamma-ray activity is found during the neutrino detection period, probably implying a large opacity for GeV gamma rays in the neutrino production region. {In our model, continuous particle acceleration/injection takes place in the inner blob at the jet base, where the hot X-ray corona of the supermassive black hole provides target photon fields for efficient neutrino production and strong GeV gamma-ray absorption. We show that this model can self-consistently interpret the neutrino emission from both two blazars in a large parameter space. In the meantime, the dissipation processes in outer blob are responsible for the simultaneous multi-wavelength emission of both sources.} In agreement with previous studies of TXS 0506+056 and, an intense MeV emission from the induced electromagnetic cascade in the inner blob is robustly expected to accompany the neutrino flare in our model could be used to test the model with the next-generation MeV gamma-ray detector in the future.
\end{abstract}


\section{Introduction}\label{intro}
The origin of extragalactic high-energy neutrinos detected by IceCube Neutrino Observatory is still unclear \citep{2013PhRvL.111b1103A, 2013Sci...342E...1I}.  As one of the most powerful astrophysical persistent objects, blazars are widely considered as source candidates for the origin of extragalactic high-energy cosmic rays and neutrinos \citep[e.g.,][]{1992A&A...260L...1M, 2001PhRvL..87v1102A, 2014PhRvD..90b3007M, 2015MNRAS.452.1877P, 2015MNRAS.448.2412P, 2016MNRAS.457.3582P}. Although stacking analysis done by \cite{2017ApJ...835...45A} suggests that blazars cannot account for the dominant diffuse neutrino background \citep[see also][but see \cite{2019ApJ...871...41P}]{2020PhRvD.101j3015L}, some individual blazars are still likely to be powerful high-energy cosmic-ray accelerators and emit high-energy neutrinos \citep{2016NatPh..12..807K, 2018Sci...361.1378I, 2019ApJ...880..103G, 2019MNRAS.489.4347O, 2020A&A...640L...4G}. 

A breakthrough in neutrino astronomy was made by IceCube in 2017. For the first time, a high-energy ($290\,\rm TeV$) muon neutrino event, IceCube-170922A, was detected in both spatial and temporal coincidence with the $\gamma$-ray flare of the known blazar TXS~0506+056 \citep{2018Sci...361.1378I} at the significance level of $\sim 3\sigma$, triggering extensive studies on the neutrino--blazar association \citep{2018ApJ...863L..10A,2018ApJ...864...84K, 2018MNRAS.480..192P, 2018ApJ...866..109S, 2018arXiv180711069Z, 2019PhRvD..99j3006B, 2019MNRAS.483L..12C, 2019NatAs...3...88G, 2019PhRvD.100j3002L, 2019PhRvD..99f3008L, 2019MNRAS.484L.104P, 2019MNRAS.483L.127R, 2019ApJ...886...23X, 2020PASJ...72...20C}. More interestingly, \citet{2018Sci...361..147I} investigated the archive data of IceCube in the direction of the same blazar, and discovered a 3.5$\sigma$ excess of 13$\pm$5 high-energy neutrino events, also known as 2014-2015 (hereafter 14-15) neutrino flare, in the period between 2014 September and 2015 March. It is worth noting that no evidence of multi-wavelength activity is found during that time (i.e., ``orphan'' neutrino flare), and the neutrino flux is about 5 times higher than the average $\gamma$-ray flux, probably implying a strong absorption of GeV photons by intense X-ray radiation field \citep{2018arXiv180900601W, 2019ApJ...881...46R}.  Such a phenomenon poses a great challenge to the conventional one-zone models. The main difficulty lies in reconciling the comparatively high flux of neutrinos and comparatively low flux of multi-wavelength electromagnetic (EM) emissions, because the emission of the EM cascade which accompanies neutrino emission, would easily overproduce the X-ray flux or the GeV gamma-ray flux of the blazar during the neutrino flare. Indeed, \cite{2019ApJ...874L..29R} study external photon fields in the broad-line region (including broad emission lines and isotropized accretion disk radiation) as the possible targets, and they found that the derived neutrino flux is not sufficient to explain the neutrino flare on the premise that the model does not overpredict the multiwavelength flux; \citet{2020ApJ...891..115P} speculated that the non-thermal photons from the layer of a structured jet could be possible targets although a detailed modelling is yet to be further explored.

Given that the conventional one-zone model is only an approximate description to the radiation processes of a parsec-scale (or even larger) jet residing in the complex environment of the very central region of the galaxy hosting an active supermassive black hole (SMBH), it may not contain all the physical processes relevant with the blazar's emission. If the association between the neutrino flare and the blazar is true, additional physical processes and/or new physical pictures have to be introduced. In fact, efforts have already been made to explore possible explanations to the neutrino flare beyond the framework of the conventional one-zone model. The jet-cloud/star model \citep{2018arXiv180900601W} and the neutral beam model \citep{2020ApJ...889..118Z} provide possible solutions but special or extreme conditions need to be satisfied. An additional compact core with enhanced activity inside a conventional emitting blob is considered \citep{2019ApJ...874L..29R} but the resulting neutrino flux is still insufficient. On the other hand, \citet{2019PhRvD..99f3008L} and \citet{2019ApJ...886...23X} considered that more than one dissipation (particle acceleration) zones could occur in the blazar's jet and form multiple radiation zones. By invoking two physically distinct emission zones in the jet, with an inner blob close to or inside of the broad-line region (BLR) and an outer blob far away from the BLR (hereafter, we denote the model by the ``inner--outer blob model''), they show that the different physical conditions of the two blobs can help suppress the EM cascade emission in the X-ray band. Consequently the theoretical neutrino flux in the model can be enhanced by at least one order of magnitude compared to that in one-zone models, when explaining the simultaneously multiwavelength emission of TXS~0506+056. A detailed comparison between the inner-outer blob model and one-zone models is given in \cite{2019ApJ...886...23X}. In the meantime, a rough analytical estimation given in \citet{2019ApJ...886...23X} suggests that the ``inner--outer'' blob model may have the potential to explain the 14-15 neutrino flare as well. If the inner blob happens to occur at the jet base, the intense X-ray photons from hot corona of the accretion disk of the SMBH could be the targets for the neutrino production and also absorb the GeV photons so that no gamma-ray flare would be accompanied with the neutrino flare. 

More recently, the second possible association between a blazar, 4FGL~J0955.1+3551, and a muon neutrino event, IceCube-200107A, is reported \citep{2020A&A...640L...4G, 2020arXiv200306012P}. 4FGL~J0955.1+3551 is a high-synchrotron peaked \citep[HSP;][]{2010ApJ...716...30A} BL Lac object at redshift $z=0.557$ \citep{2020MNRAS.tmpL..59P}. Its X-ray flux was found in a high state with a factor of 2.5 larger than the average flux in 2012-2013 when \textit{Swift} started the follow-up observation the day after the arrival of IceCube-200107A \citep{2020ATel13394....1G}. Assuming the spectral index of neutrino spectrum to be $-2$ ($N_{\nu}(E)\propto E^{-2}$) and the effective area for HESE starting tracks \citep{2019ICRC...36.1021B}, \cite{2020A&A...640L...4G} calculate that the all-flavour neutrino energy flux between 65~TeV and 2.6~PeV for 30 days, 250 days and 10 years are $3\times 10^{-9}~\rm erg~cm^{-2}~s^{-1}$, $4\times 10^{-10}~\rm erg~cm^{-2}~s^{-1}$ and $3\times 10^{-11}~\rm erg~cm^{-2}~s^{-1}$, respectively. The GeV gamma-ray flux is at the level of $10^{-12}\rm erg~cm^{-2}s^{-1}$ which is more than one order of magnitude lower than the neutrino flux even assuming a 10-yr emission period. \cite{2020A&A...640L...4G} argue that IceCube-200107A can only be interpreted as an upward fluctuation  \citep[see also][]{2020ApJ...899..113P} or a random event detected from a numerous population of faint sources (i.e., the Eddington bias, \citealt{2019A&A...622L...9S}) in the conventional one-zone models, given the huge contrast between the neutrino flux and the GeV gamma-ray flux. Moreover, with analytical estimations, \cite{2020arXiv200306012P} conclude that any theoretical model on explaining IceCube-200107A from 4FGL~J0955.1+3551 must involve an intense external photon field.

In this work, we aim to explore that whether the 14-15 neutrino flare and IceCube-200107A could physically originate, respectively, from TXS~0506+056 and 4FGL~J0955.1+3551 under the inner--outer blob model, without ascribing the detection to the Eddington bias. The rest of this paper is organized as follows. In Section \ref{MD}, we give a brief description of the model and the setup in this work; in Section \ref{results}, we apply the model to explain the multi-messenger emission of 14-15 neutrino flare from TXS~0506+056 and IceCube-200107A from 4FGL~J0955.1+3551 and perform an extensive investigation on the parameter space of the inner blob based on the constraints from the EM observations; in Section \ref{DC}, we discuss our results and give a conclusion. Throughout the paper, the $\Lambda$CDM cosmology with $H_{\rm{0}}={70\rm{km~s^{-1}~Mpc^{-1}}}$, $\Omega_{\rm{m}}=0.3$, $\Omega_{\rm{\Lambda}}=0.7$ is adopted.

\begin{figure*}[htbp]
\centering
\includegraphics[width=0.6\textwidth]{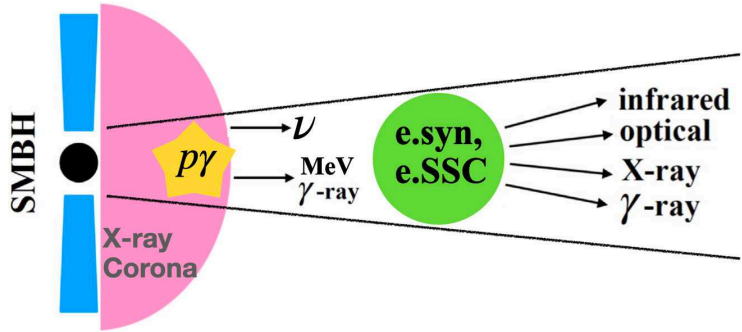}
\caption{A schematic illustration (not to scale) of the inner--outer blob model. The outer blob (green circle) is far away from the BLR and the inner blob (yellow star) is near or at the jet base with a distance comparable to only a few times larger than the Schwarzschild radius of the SMBH. The external photons from hot corona would  provide copious targets for hadronic interactions of protons as well as the EC scattering of electrons. On the other hand, the intensive X-ray photons from hot corona also absorb the GeV-TeV gamma rays and generate significant cascades emission, with energy around MeV band. The low-energy synchrotron radiation is suppressed due to the dominant Compton cooling of electrons. The observed multi-wavelength radiation arises from the synchrotron emission and SSC emission of electrons (e.syn and e.SSC) from the outer blob where the EC radiation is negligible.
\label{fig:sketch}}
\end{figure*}

\section{Model Description}\label{MD}
Let us begin with a brief description about the physical picture of the inner--outer blob model and its basic assumptions. Henceforth, physical quantities with the superscript “AGN” are measured in the AGN frame, whereas quantities without the superscript are measured in the jet's comoving frame, unless specified otherwise. 

Following the setup in \cite{2019ApJ...886...23X}, two emitting blobs in the jet, i.e., inner blob and outer blob, are considered in the modeling. We assume that these two blobs are spherical plasmoid of different radii ($R_{\rm in}<R_{\rm out}$), moving with the same bulk Lorentz factor $\Gamma_j$ along the jet axis. Assuming the observer view the blob at an angle $\theta$ with respect to the jet axis, the Doppler factor of the two blobs are $\delta_{\rm D}=1/\Gamma_j(1-\beta_j\cos\theta)$\footnote{For simplicity, we assume the Doppler factor $\delta_{\rm D} \approx \Gamma_j$ for a relativistic jet close to the line of sight in blazars with a viewing angle of $\theta \lesssim 1/\Gamma$ hereafter.}. The two blobs are filled with uniformly entangled magnetic fields ($B_{\rm in}>B_{\rm out}$), relativistic electrons and protons. Electrons and protons are assumed to be injected into both blobs with a broken power-law distribution 
\begin{equation}\label{eq:einj}
\begin{split}
Q_{\rm e}(\gamma_{\rm e}) = &Q_{\rm e, 0}\gamma_{\rm e}^{-s_{\rm e, 1}}\left[1 + \left(\frac{\gamma_{\rm e}}{\gamma_{\rm e, b}}\right)^{(s_{\rm e, 2} - s_{\rm e, 1})}\right]^{-1}, \\
&{\rm for} ~~ \gamma_{\rm e, min} < \gamma_{\rm e} < \gamma_{\rm e, max},
\end{split}
\end{equation}
and a power-law distribution
\begin{equation}\label{eq:pinj}
Q_{\rm p}(\gamma_{\rm p}) = Q_{\rm p, 0}\gamma_{\rm p}^{-s_{\rm p}}, \gamma_{\rm p, min} < \gamma_{\rm p} < \gamma_{\rm p, max},
\end{equation}
respectively. The free parameters for the spectral shape of electrons are the minimum, break, and maximum Lorentz factors $\gamma_{\rm e, min/b/max}$ and the two spectral indices $s_{\rm e, 1/2}$, respectively. For the spectral shape of protons, the free parameters are the minimum, and maximum Lorentz factors $\gamma_{\rm p, min/max}$ and the spectral index $s_{\rm p}$, respectively.

\begin{figure}[htbp]
\centering
\includegraphics[width=1\columnwidth]{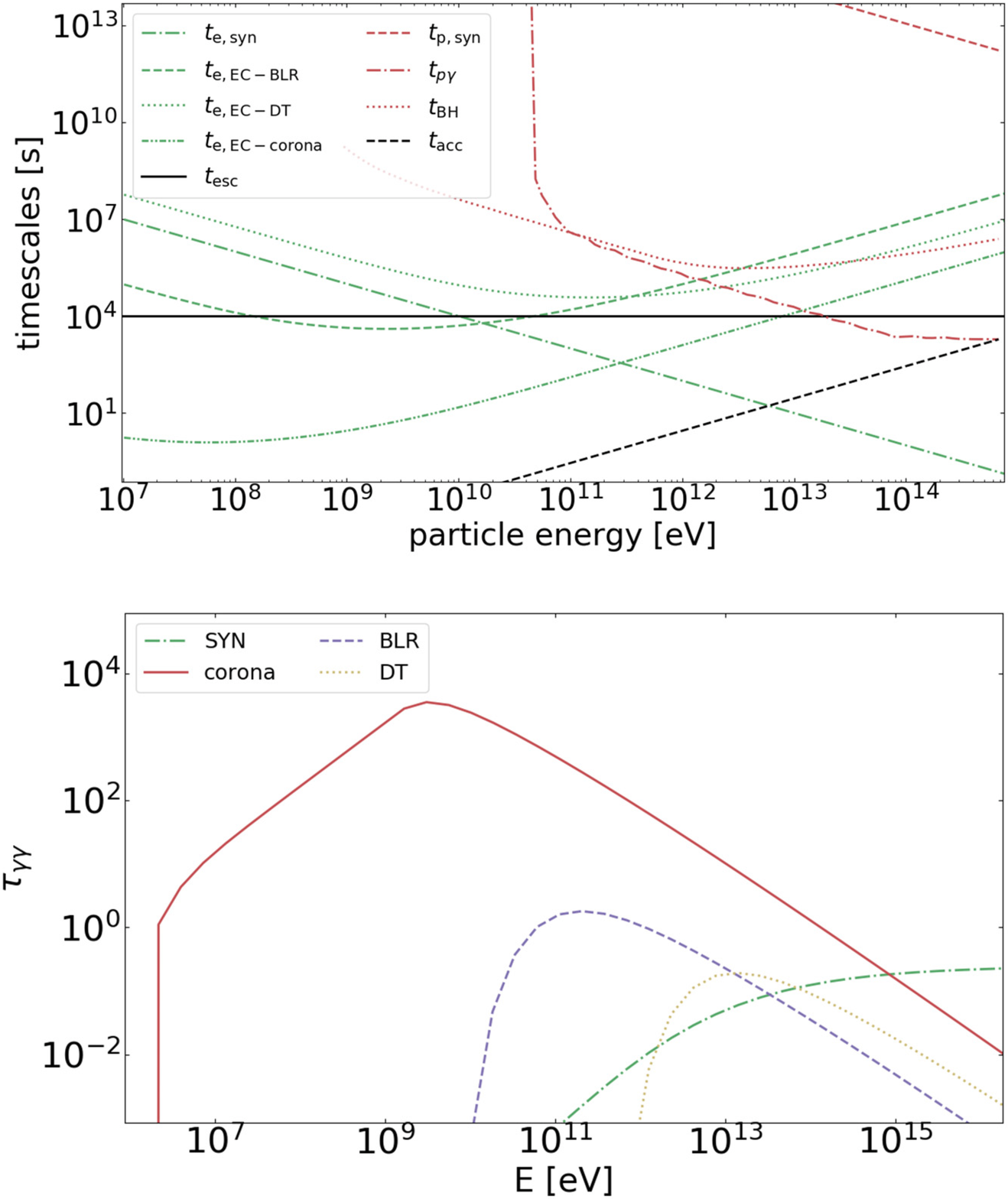}
\caption{Upper panel: Timescales of various cooling processes for electrons (green curves) and protons (red curves) in the inner blob of TXS~0506+056 as a function of the particle energy. Both the particle energy and timescale are measured in the jet comoving frame. The parameters are the same as those used in Table~\ref{tab1}. The black horizontal lines denote the escape timescale of particles in the ballistic propagation limit. The meaning of all curves is explained in the inset legend. Lower panel: Comparison of the opacity for $\gamma\gamma$ annihilation as a function of the photon energy in the observer's frame that contributed by the photons from synchrotron emission of primary electrons (dot-dashed green curve), hot corona (solid red curve), broad-line region (dashed purple curve) and dusty torus (dotted yellow curve), respectively.
\label{fig:time}}
\end{figure}

%
The outer blob is assumed to be far away from the BLR, therefore the energy density of external photons in the outer blob is so weak that it can be neglected in the corresponding leptonic and hadronic processes. Therefore, the hadronic emission can be neglected
and relativistic electrons radiate mainly through synchrotron radiation and synchrotron self-Compton (SSC) scatterings, giving rise to multi-wavelength emission. The inner blob is closer to the SMBH, so the radiation of the BLR and the dusty torus (DT) could enter the blob with a significant amplification of the energy density due to the Doppler effect. Those external radiation fields provide copious targets for hadronic interactions of protons as well as the external Compton (EC) scattering of electrons (see the upper panel of Fig.~\ref{fig:time}). The former process gives rise to neutrino emission and the latter process is relevant for gamma-ray emission. The energy density of BLR ($u_{\rm BLR}$) and DT ($u_{\rm DT}$) in the comoving frame as a function of the distance along the jet, $r^{\rm AGN}_{\rm in}$, can be approximately written as \citep{2012ApJ...754..114H}:
\begin{equation}\label{blr}
u_{\rm BLR} \approx \frac{\phi_{\rm BLR} \Gamma_j^2L_{\rm d}}{4\pi (r_{\rm BLR}^{\rm AGN})^2c[1+(r^{\rm AGN}_{\rm in}/r_{\rm BLR}^{\rm AGN})^3]},
\end{equation}
and
\begin{equation}\label{dt}
u_{\rm DT} \approx \frac{\phi_{\rm DT} \Gamma_j^2L_{\rm d}}{4\pi (r_{\rm DT}^{\rm AGN})^2c[1+(r^{\rm AGN}_{\rm in}/r_{\rm DT}^{\rm AGN})^4]},
\end{equation}
where $\phi_{\rm BLR}=\phi_{\rm DT}=0.1$ are the fractions of the disk luminosity $L_{\rm d}$ reprocessed into the BLR and DT radiation, respectively. It is assumed that the characteristic distance of BLR in the AGN frame is $r_{\rm BLR}^{\rm AGN} = 0.1(L_{\rm d}/10^{46}\rm erg \,  s^{-1})^{1/2}$~pc and the characteristic distance of the DT is $r_{\rm DT}^{\rm AGN} = 2.5(L_{\rm d}/10^{46}\rm erg \,  s^{-1})^{1/2}$~pc \citep[e.g.,][]{2008MNRAS.387.1669G}. The radiation from both the BLR and DT is taken as an isotropic blackbody with a peak at $2\times10^{15}\Gamma_j$ Hz \citep{2008MNRAS.386..945T} and $3\times10^{13}\Gamma_j$ Hz \citep{2007ApJ...660..117C} in the comoving frame. 
In addition, since no gamma-ray activities are detected by \textit{Fermi}-LAT for both 14-15 neutrino flare and IceCube-200107A, GeV gamma-ray photons must be absorbed, implying a compact size for the inner blob and an environment immersed in the intense X-ray radiation. Here, we consider that the inner blob is near or at the jet base with a distance comparable to only a few times larger than the Schwarzschild radius of the central SMBH ($r_{\rm Sch}\sim (M_{\rm SMBH}/5\times10^8M_\odot)\times10^{14}~\rm cm$). Then the X-ray photons emitted by the hypothetical corona surrounding the accretion disk \citep{2014ARA&A..52..589H} would enter the inner blob and interact with the gamma-ray photons \citep{2019MNRAS.483L.127R}. The emission of the hot corona typically has a power-law spectrum \citep{2018ApJ...866..124K, 2018MNRAS.480.1819R}
\begin{equation}
L(E) = L_{\rm 1keV}(E/\rm 1~keV)^{1-\alpha},~~~0.1~\rm keV <\textit{E}<100~\rm keV,
\end{equation}
and the corresponding energy density in the comoving frame can be estimated as
\begin{equation}
u_{\rm corona}=\frac{\Gamma_j^2\int L(E)/E \rm{d}\it{E}}{4\pi {r_{\rm in}^{\rm AGN}}^2c}.
\end{equation}
Note that the size of the hot corona is generally only several times Schwarzschild radius, so the time that a relativistically moving blob need to pass through the hot corona, as measured by an observer, would be only $\sim r_{\rm Sch}/c\Gamma_j\delta_{\rm D}\sim 10^2\,$s which is much shorter than the duration of the 14-15 neutrino flare or the considered emission period of IceCube-200107A in the previous literature. Therefore, multiple blobs need to form in the hot corona region consecutively, or alternatively the inner blob could be related to certain dissipation process that occurs at a quasi-stationary feature such as a standing shock at the bottom of the jet \citep{2011A&A...531A..95F}, so that the neutrino radiation in the inner blob can last a period comparable to the observation. As a consequence, the inner blob would appear as a quasi-stationary emitting zone, and the Doppler boosting is different from that for the outer blob. For the inner blob, we have $F_{\nu}^{\rm obs}\propto \delta_{\rm D}^2 F_{\nu}^{\rm in}$ \citep[e.g.,][]{2009arXiv0909.2576M, 2020PhRvD.102h3028L}, while for the outer blob, there is another factor of $\delta_{\rm D}$ from the time compression since the entire emission region is moving towards us, then we have $F_{\nu}^{\rm obs}\propto \delta_{\rm D}^3 F_{\nu}^{\rm out}$. A sketch of the physical picture of our model is shown in Fig.~\ref{fig:sketch}. 

To avoid introducing too many free parameters, the strategies for reducing the free parameters in \cite{2019ApJ...886...23X} are applied here. 
\begin{enumerate}
\item We assume the two blobs have the same spectral shape for the injected particles, the same electron injection luminosity $L_{\rm e, inj}$, and the same proton injection luminosity $L_{\rm p, inj}$.

\item We set the BLR luminosities $L_{\rm BLR}^{\rm AGN}$ of TXS~0506+056 and 4FGL~J0955.1+3551 are both $5\times 10^{43}~\rm erg/s$, as estimated by \cite{2019MNRAS.484L.104P} and \cite{2020arXiv200306012P}. Then the disk luminosity $L_{\rm d}$ can be estimated through $L_{\rm d}=10\times L_{\rm BLR}^{\rm AGN}$ as proposed by \cite{2008MNRAS.387.1669G}.

\item Since the X-ray radiation from hot corona of blazars is not directly observed, here we simply assume that its spectral index $\alpha=1$, which is the observational typical value of AGN \citep{2013peag.book.....N} and $L_{\rm 1keV}$ is comparable to BLR luminosity. Thus, $L_{\rm 1keV}$ is set as $5\times 10^{43}~\rm erg/s$ for the two objects.

\item Since we already assume the external photons from BLR and DT are not relevant for the leptonic and hadronic processes, we do not specify the location of the outer blob $r_{\rm out}^{\rm AGN}$. For the inner blob, its distance from SMBH $r_{\rm in}^{\rm AGN}$ is set to be a free parameter but restricted to be a few times of the Schwarzschild radius $r_{\rm Sch}$. 

\item Since the inner blob is located at the jet base, the radius of the inner blob should be $R_{\rm in}\leq r_{\rm in}^{\rm AGN}$. On the other hand, the external radiation field is the dominant targets for the hadronic interactions and EC scatterings, and its energy density is only related to position of inner blob $r_{\rm in}^{\rm AGN}$, therefore the specific value of $R_{\rm in}$ will not significantly change the fitting results. For simplicity, we assume $R_{\rm in}=r_{\rm in}^{\rm AGN}$\footnote{Radio observations of some objects \citep[e.g.,][]{2012ApJ...745L..28A} suggest that AGN jets may have a parabolic base, i.e., the jet is not collimated at the base, which may validate this assumption.}.

\item For the minimum and maximum electron Lorentz factors, we set $\gamma_{\rm e, min}=50$ and $\gamma_{\rm e, max}=10^7$, since our modeling results are not sensitive to them.

\item The minimum proton Lorentz factor $\gamma_{\rm p, min}$ is fixed to be 1 and we set spectral index $s_{\rm p}$ of proton energy distribution at injection to be 2.

\item $\gamma_{\rm p, max}$ is obtained by equating the acceleration and the cooling or escape time-scales
\begin{equation}
t_{\rm acc} = \min\{ t_{p\gamma}, t_{\rm BH}, t_{\rm esc}, t_{\rm p, syn}\}.
\end{equation}
If we assume the particle acceleration is dominated by Fermi-type acceleration \citep[e.g.,][]{2007Ap&SS.309..119R}, the acceleration time-scale $t_{\rm acc}$ can be expressed as $t_{\rm acc}=\eta \gamma_{\rm p}m_{\rm p}c/eB_{\rm in} = 2.6\times10^{-3}\gamma_{\rm p}(\frac{\eta}{50})(\frac{B_{\rm in}}{2~\rm G})^{-1}$, where $\eta$ is an efficiency factor characterizing the acceleration rate. $t_{p\gamma} \simeq [c n_{\rm soft}<\sigma_{p\gamma}\kappa_{p\gamma}>]^{-1}$ is the $p\gamma$ energy loss time-scale, where $n_{\rm soft}$ is the number density of the soft photons and $<\sigma_{p\gamma}\kappa_{p\gamma}>\simeq10^{-28} \rm cm^2$ is the inelasticity-weighted $p\gamma$ interaction cross-section. $t_{\rm BH} \simeq [c n_{\rm soft}<\sigma_{\rm BH}\kappa_{\rm BH}>]^{-1}$ is the BH pair-production cooling time-scale, where $\sigma_{\rm BH}$ and $\kappa_{\rm BH}$ are the cross section and inelasticity for the BH pair production process \citep{1992ApJ...400..181C}. $t_{\rm esc}=R_{\rm in}/c$ is the escape time-scale. $t_{\rm p, syn} = 6\pi m_{\rm e}c^2/(c\sigma_{\rm T}B^2\gamma_{\rm p})(m_{\rm p}/m_{\rm e})^3$ is the proton-synchrotron cooling timescale. It can be seen from the upper panel of Fig.~\ref{fig:time} that due to the very high energy density of the external photons (especially the photons from hot corona), the $p\gamma$ energy loss above tens TeV is very efficient, dominating over escape. Therefore, under adopted parameters in Fig.~\ref{fig:time}, $\gamma_{\rm p, max}$ is determined by the equality $t_{\rm acc} = t_{p\gamma}$.
\end{enumerate}
At last, the number of free parameters is reduced to eleven, namely $\delta_{\rm D}$, $B_{\rm in}$, $B_{\rm out}$, $R_{\rm in}$, $R_{\rm out}$, $L_{\rm e, inj}$, $L_{\rm p, inj}$, $s_{\rm e, 1}$, $s_{\rm e, 2}$, $\gamma_{\rm e, b}$, and $\eta$. 

With the above setups, the synchrotron, SSC and EC photons from primary electrons, and external photons from hot corona, BLR and DT are all considered as the targets for $p\gamma$ interactions and $\gamma\gamma$ pair-production in the inner blob. The photopion production and Bethe-Heitler (BH) pair-production processes are calculated following \cite{2008PhRvD..78c4013K}. Triggered by the photons from neutral pion decay, secondary electron/positron pairs from charged pion decay, and BH pair-production, the Compton-supported cascades in the inner blob are evaluated using a semi-analytical method developed by \cite{2013ApJ...768...54B} and \cite{2018ApJ...857...24W}. Since the very high-energy gamma-ray photons escape from the inner or the outer blobs will be absorbed by the extragalactic background light (EBL), we calculate the attenuation in the GeV-TeV band using the EBL model of \cite{2011MNRAS.410.2556D}.

\begin{table*}
\caption{Model parameters for SED fit of TXS~0506+056 shown in the upper panel of Fig.~\ref{fig:sed}\label{tab1}}
\centering
\begin{tabular}{@{}llllllllllll@{}}
\hline\hline
Free parameters&$\delta_{\rm D}$&$B_{\rm out}$&$B_{\rm in}$&$R_{\rm out}$&$R_{\rm in}$&$L_{\rm e, inj}$&$L_{\rm p, inj}$&     $s_{\rm e, 1}$ & $s_{\rm e, 2}$&$\gamma_{\rm e, b}$&$\eta$\\
 & &  [G] & [G] & [cm] &  [cm] &  [erg/s] & [erg/s] & & & &  \\
 \hline
Values & 20 & 0.75 & 2 & $2\times10^{16}$ & $3\times10^{14}$ & $6\times10^{41}$  & $1.1\times10^{44}$ &  1.4 &  4 & $1.5\times10^{4}$ & 50 \\
\hline
\hline
Fixed/Derived  &  $L_{\rm BLR}^{\rm AGN}$ & $r^{\rm AGN}_{\rm in}$ & $\gamma_{\rm e, min}$ & $\gamma_{\rm e, max}$ & $s_{\rm p}$ & $\gamma_{\rm p, min}$ & $\gamma_{\rm p, max}$ & $L_{\rm e, in}^{\rm k}$ & $L_{\rm p,in}^{\rm k}$ & \\
parameters & [erg/s] &  [cm] & & & & & &[erg/s] &[erg/s] & \\
\hline
Values & $5\times10^{43}$ & $3\times10^{14}$ &  50 & $10^7$ & 2 & 1 & $7.2\times10^5$ &  $2.4\times10^{41}$ & $2.44\times10^{46}$ & \\
\hline
\end{tabular}
\tablecomments{The kinetic luminosity in relativistic electrons $L_{\rm e, in}^{\rm k}$ and in relativistic protons $L_{\rm p,in}^{\rm k}$ in the inner blob are calculated as $L_{\rm e,in}^{\rm k} \simeq \pi R_{\rm in}^2\delta_{\rm D}^2 m_{\rm e}c^3\int \gamma_{\rm e}N_{\rm e}(\gamma_{\rm e})d\gamma_{\rm e}$ and $L_{\rm p, in}^{\rm k} \simeq \pi R_{\rm in}^2\delta_{\rm D}^2 m_{\rm p}c^3\int \gamma_{\rm p}N_{\rm p}(\gamma_{\rm p})d\gamma_{\rm p}$, where $N_{\rm e}(\gamma_{\rm e})$ and $N_{\rm p}(\gamma_{\rm p})$ are the steady-state electrons and protons energy distributions, respectively.}
\end{table*}

\begin{table*}
\caption{Model parameters for SED fit of 4FGL~J0955.1+3551 shown in the lower panel of Fig.~\ref{fig:sed}\label{tab2}}
\centering
\begin{tabular}{@{}lllllllllllll@{}}
\hline\hline
Free parameters&$\delta_{\rm D}$&$B_{\rm out}$&$B_{\rm in}$&$R_{\rm out}$&$R_{\rm in}$&$L_{\rm e, inj}$&$L_{\rm p, inj}$&     $s_{\rm e, 1}$ & $s_{\rm e, 2}$&$\gamma_{\rm e, b}$&$\eta$\\
 & &  [G] & [G] & [cm] &  [cm] &  [erg/s] & [erg/s] &  & & &  \\
 \hline
Values & 20 & 0.16 & 2 & $3\times10^{16}$ & $3\times10^{14}$ & $2\times10^{41}$  & $1.2\times10^{44}$ &  1.6 &  4 & $5\times10^{5}$ & 50 \\
\hline
\hline
Fixed/Derived  &  $L_{\rm BLR}^{\rm AGN}$ & $r^{\rm AGN}_{\rm in}$ & $\gamma_{\rm e, min}$ & $\gamma_{\rm e, max}$ & $s_{\rm p}$ & $\gamma_{\rm p, min}$ & $\gamma_{\rm p, max}$ & $L_{\rm e, in}^{\rm k}$ & $L_{\rm p,in}^{\rm k}$ & \\
parameters & [erg/s] & [cm] & & & & &  &[erg/s] &[erg/s] & \\
\hline
Values & $5\times10^{43}$ & $3\times10^{14}$ & 50 & $10^7$ & 2 & 1 & $8.3\times10^5$ &  $4.6\times10^{41}$ & $2.6\times10^{46}$ & \\
\hline
\end{tabular}
\end{table*}

\begin{figure}[htbp]
\centering
\includegraphics[width=1\columnwidth]{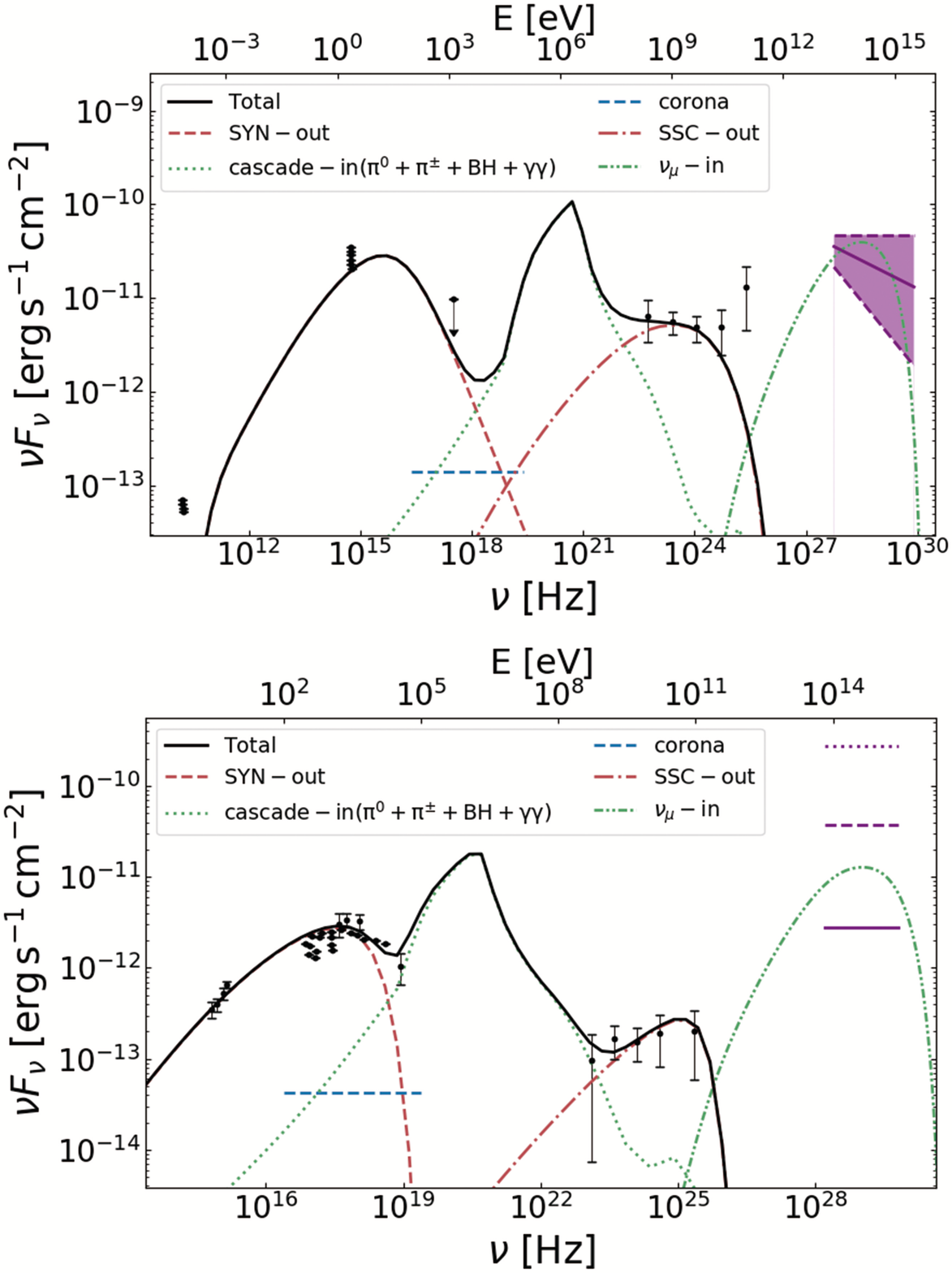}
\caption{Upper panel: The multi-messenger emissions of TXS~0506+056 predicted by the two-zone model. The quasi-simultaneous data is taken from \cite{2018MNRAS.480..192P} and the violet bow tie shows the muon-neutrino flux as measured during 14-15 flare \citep{2018Sci...361..147I}. Lower panel: The multi-messenger emissions of 4FGL~J0955.1+3551 predicted by the two-zone model. The quasi-simultaneous data is taken from \cite{2020arXiv200306012P}, and the violet dotted, dashed and solid horizontal lines show the $\nu_\mu+\bar{\nu}_\mu$ neutrino flux for 30 days, 250 days and 10 years \citep{2020A&A...640L...4G}. The line styles in both panels have the same meaning. The dashed red and dot-dashed red curves represent the synchrotron and SSC emission from primary electrons in the outer blob, respectively. The dotted green curve represents the emission from pair cascades in the inner blob. The double dot-dashed green curve shows the muon neutrino energy spectrum. The solid black curve is the total emission from both blobs.
\label{fig:sed}}
\end{figure}

\section{Results}\label{results}

Observationally,  during the 14-15 neutrino flare, no EM activity in any wavelength is observed from TXS~0506+056. This feature constrains that the EM emission from neutrino production region can not contribute significantly to the observed energy band. The UV/X-ray emission of 4FGL~J0955.1+3551 is found in the high-state at the time of the arrival of IceCube-200107A. However, no activity is found in other wavelengths. Besides, if the neutrino emission period is assumed to be 250\,days or 10\,years \citep{2020A&A...640L...4G}, the high-state X-ray emission may not be causally linked to the neutrino emission. We therefore speculate that IceCube-200107A is produced in a similar process to that for the 14-15 neutrino flare of TXS~0506+056. 
In the modeling, we find that there are many sets of parameters may satisfactorily reproduce the multi-messenger observations of TXS~0506+056 and 4FGL~J0955.1+3551. Here we firstly show the fitting results with a specific set of parameters for TXS~0506+056 and 4FGL~J0955.1+3551, respectively,  as shown in Fig.~\ref{fig:sed}. The corresponding parameters are shown in Table~\ref{tab1} and Table~\ref{tab2}. From Fig.~\ref{fig:sed}, it can be seen that, as expected, the multi-wavelength SEDs \textit{with measurements} are dominated by the synchrotron and SSC emission emitted by the primary electrons from the outer blob, which do not emit high-energy neutrinos, while the neutrinos originate from the inner blob, accompanied by a comparable flux in the MeV band. The high MeV flux is a consequence of the EM cascades developed in the intense X-ray radiation field from the hot corona {(see also the lower panel of Fig.~\ref{fig:time})}: given that the X-ray spectrum of the hot corona can extend up to $\sim 100$\,keV, it can absorb $\gamma$-ray photons starting from $E_{\gamma,\rm th}\approx \frac{2(m_{\rm e}c^2)^2}{100~\rm keV(1+\it{z})^2}\simeq 2\,$MeV in the observer's frame with an opacity at 2\,MeV being $\tau_{\gamma\gamma}=\frac{\sigma_{\gamma\gamma}L_{\rm 100~keV}}{4\pi r_{\rm in}^{\rm AGN}c100~\rm keV}\simeq 0.5$. The high-energy gamma-ray photons and electron/positron pairs generated in the $p\gamma$ interactions will initiate EM cascades and generate lower and lower energy photons or electrons/positrons with the development of the cascade, until the energy of the newly generated photons drop close to $E_{\gamma,\rm th}$. These photons will be absorbed by the X-ray radiation field and give birth to the last-generation electron/positron pairs of the EM cascade, with energy $E_{\gamma,\rm th}/2\gtrsim 1\,$MeV. Therefore, most energies of the first-generation gamma rays and electron/positron pairs that co-produced with neutrinos will deposit to the pairs of $\gtrsim 1$\,MeV, which mainly radiate in the MeV band via the IC scattering off the X-ray photons of the hot corona because the energy density of hot corona dominate over the energy densities of magnetic field and other photon fields\footnote{In the modeling, $r_{\rm in}^{\rm AGN}/r_{\rm BLR}^{\rm AGN}\approx 4\times10^{-3}$, therefore $u_{\rm corona}/u_{\rm BLR}\approx (r_{\rm in}^{\rm AGN}/r_{\rm BLR}^{\rm AGN})^2\sim 10^5$. The ratio of $u_{\rm corona}$ to $u_{\rm DT}$ can be estimated in the same way, which is $\sim 10^8$.} \citep[see also Figure 10 of][]{2019ApJ...881...46R}. Therefore, in the absence of measurements at MeV band, it seems as if the blazars underwent ``orphan'' neutrino flares. 

From the upper panel of Fig.~\ref{fig:sed}, we can see that the multi-wavelength SED of TXS 0506+056 is well fitted by the leptonic emission from outer blob except the highest energy GeV data point which shows a spectral hardening, although a hardening feature may not be significant \citep{2019ApJ...880..103G}. {This is because the inner blob has a large gamma-ray opacity while the outer blob, which is far away from the BLR and DT radiation field, can hardly produce a hard spectrum extending up to 10\,GeV (or above) solely with the SSC process. This is the limiation of the present model. However, in the physical picture of the inner-outer model, there could be additional dissipative regions being formed between the inner and outer blobs (i.e., at a distance $\sim r_{\rm BLR}^{\rm AGN}$ from the SMBH), which can in principle reproduce the highest-energy GeV data point with the EC process. We refrain ourselves from the further discussion on the interpretation of the last GeV data point and focus back on the neutrino flare.} We can see that the inferred neutrino flux during the 14-15 flare of TXS 0506+056 can be well reproduced by our model. For 4FGL~J0955.1+3551, the predicted neutrino flux is much higher that of one-zone model \citep{2020arXiv200306012P}. This is because of the differences between the one-zone model and the inner-outer blob model. As indicated in \cite{2019ApJ...886...23X}, by ascribing the low-energy emission (radio to UV band) of the blazar to the outer blob, synchrotron emissivity of primary electrons in the inner blob could be much lower than that in the one-zone models, especially in the presence of dense external photon fields. As a result, the $\gamma \gamma$ absorption opacity for sub-PeV/PeV photons is significantly reduced and less than unity (see the lower panel of Fig.\ref{fig:time}) in the inner blob. This suppresses the cascade emission by reducing the number of secondary electrons generated in $\gamma \gamma$ absorption. Also, the secondary electrons generated in the cascade preferentially emit MeV-GeV gamma rays through the IC scattering in the dense external photon fields rather than in the X-ray band through the synchrotron radiation, and hence the restrictive constraints by the X-ray observation is avoided. In our modeling, the predicted neutrino flux is well above the one assuming an emission period of 10\,years given by \citet{2020A&A...640L...4G}. As shown in Fig.~\ref{fig:time}, since the energy density of the hot corona is very high, the $p\gamma$ interaction is very efficient ($f_{p\gamma}=t_{\rm esc}/t_{p\gamma} \approx 3$ for the tens TeV to PeV protons). Therefore, the required proton injection luminosity in the inner-outer blob model is moderate and reasonable from the point of view of energy budget. In our modeling, the baryon loading factors, defined as $L_{\rm p, inj}/L_{\rm e, inj}$, of TXS 0506+056 and 4FGL J0955.1+3551 are 125 and 600, respectively, which are significant lower than that of the conventional one-zone proton synchrotron model \citep[e.g.,][]{2013ApJ...768...54B} and one-zone $p\gamma$ model \citep[e.g.,][]{2019ApJ...871...81X}. Provided the mass of the SMBH being $3\times 10^8M_\odot$\citep{2019MNRAS.484L.104P} , the jet's kinetic luminosity employed here is also below the Eddington limit.

\begin{figure}[htbp]
\centering
\includegraphics[width=1\columnwidth]{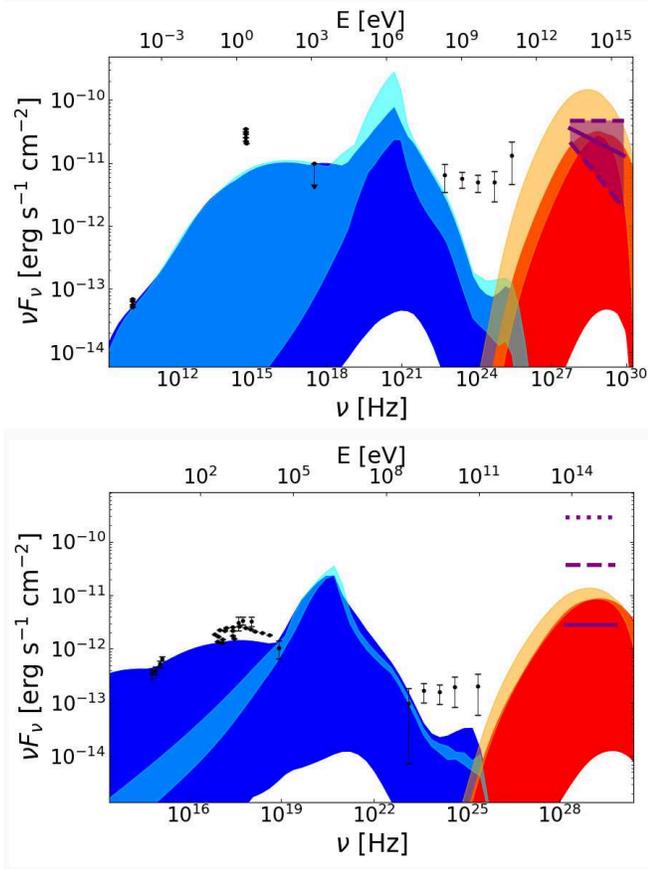}
\caption{The results of parameter scanning for the EM and neutrino emission from inner blobs of TXS~0506+056 (upper panel) and 4FGL~J0955.1+3551 (lower panel), respectively. The data points, violet bow tie and horizontal lines are the same as in Fig.~\ref{fig:sed}. The yellow/red solid {bands} represent the muon neutrino energy spectra with different parameter sets in the inner blob, which can/cannot produce a sufficient flux to explain the neutrino measurements, while the cyan/blue solid {bands} represent the corresponding emission from pair cascades. The parameter resulting in the highest (lowest) neutrino flux for TXS 0506+056 are $\delta_{\rm D}=30(10)$, $r_{\rm in}^{\rm AGN}=3\times10^{14}(10^{15})~\rm cm$, $B_{\rm in}=1(100)~\rm G$, $\eta=100(10)$, and $\delta_{\rm D}^2L_{\rm p,inj}=10^{47}(10^{45})~\rm erg/s$; the parameters resulting in the highest (lowest) neutrino flux for 4FGL~J0955.1+3551 are $\delta_{\rm D}=30(10)$, $r_{\rm in}^{\rm AGN}=3\times 10^{14}(10^{15})~\rm cm$, $B_{\rm in}=1(100)~\rm G$, $\eta=100(10)$, and $\delta_{\rm D}^2L_{\rm p,inj}=10^{46.5}(10^{45})~\rm erg/s$.
\label{fig:search}}
\end{figure}

To further explore the general applicability of the inner--outer blob model in explaining the blazar ``orphan'' neutrino flare, we perform an extensive scan of the parameter space for the two sources separately. Since both these two emissions are relevant with  the inner blob, we only search the parameter space of the inner blob. In addition, leptonic emissions from the inner blob do not contribute to the observed SED, and hence is neglected here. We scan the parameter space of five parameters for the properties of the inner blob and injected relativistic protons energy distribution within physically plausible ranges, which are $10\leqslant \delta_{\rm D}\leqslant 30$, $3\times10^{14} \leqslant r^{\rm AGN}_{\rm in}/\rm cm\leqslant 3\times10^{15}$, $1 \leqslant B_{\rm in}/\rm G\leqslant 100$, $10\leqslant \eta \leqslant100$, and $10^{45} \leqslant \delta_{\rm D}^2L_{\rm p, inj}/\rm erg~s^{-1}\leqslant 10^{47}$. The former two parameters are spaced into five bins linearly while the latter three are spaced into five bins logarithmically. Since the observed SEDs are ascribed to the emissions from the outer blob, the flux of the cascade emission should be limited below 1~$\sigma$ upward error bar of the data points. Generally speaking, a larger $\delta_{\rm D}$ and a smaller $r^{\rm AGN}_{\rm in}$ would increase the energy density of the external photon field and improve the $p\gamma$ interaction efficiency $f_{p\gamma}$. A smaller $B_{\rm in}$ would reduce the predicted X-ray flux through the synchrotron radiation of secondary pairs co-produced with neutrinos, making the result less constrained by the X-ray data and leaving more room for the neutrino production.

The scanning results are shown in Fig.~\ref{fig:search}.  For the 14-15 neutrino flare of TXS~0506+056, since the observationally inferred neutrino events is $13\pm 5$ during about 5 months, we show the neutrino flux under parameters that are able to produce at least 8 events in 10\,TeV -- 10\,PeV by yellow {band} while the red {band} show that unable to produce more than 8 events. The corresponding EM fluxes are shown in cyan and blue respectively. The result for 4FGL~J0955.1+3551 is exhibited in a similar way except coloring the {bands} based on whether the neutrino flux can give rise to 1 event detection by IceCube (yellow and cyan) or not (red and blue) in the range of 65\,TeV -- 2.6\,PeV for an emission period of 250\,days. The IceCube point-source effective area for (anti-)muon neutrinos \citep{2019ICRC...36..851C} in the declination $-5^\circ-30^\circ$ and $30^\circ-90^\circ$ are employed for TXS~0506+056 and 4FGL~J0955.1+3551 respectively to calculate the expected neutrino event\footnote{This is not to be confused with the three horizontal lines shown in in the lower panels of Fig.~\ref{fig:sed} and Fig.~\ref{fig:search}, which represent the expected neutrino flux to trigger one starting event in IceCube during the respective given emission period. The effective area for starting event is more than one order of magnitude smaller than the point-source effective area.}.  It can be seen that there are amount of parameter sets able to explain the 14-15 neutrino flare from TXS~0506+056 and IceCube-200107A from 4FGL~J0955.1+3551. For the latter, the most optimistic parameters can result in a non-negligible probability of $20\%$ for detection of one event by IceCube even assuming an  emission period of 30 days. One robust and remarkable feature in the resulting SED is the MeV flux. As is shown in Fig.~\ref{fig:search}, when the observed neutrino energy spectrum can be interpreted under certain parameter set (the yellow {band}), the corresponding cascade emission will contribute significant radiation in the MeV band (the cyan {band}). It should be also noted that a spectral bump in the MeV band is not the unique feature predicted in our model. In some other lepto-hadronic models of the 14-15 neutrino flare of TXS~0506+056, an intense MeV emission is also expected provided the presence of a dense external radiation field in the UV to X-ray band \citep[e.g.][]{2018arXiv180900601W, 2019ApJ...881...46R, 2019ApJ...874L..29R, 2020ApJ...891..115P}. On the other hand, the flux level of the MeV emission is generally comparable to the neutrino flux and might be used as an indicator of neutrino flux and distinguish different models. Currently, there is no active instruments able to measure the accompanying MeV emission during the neutrino emission period, but the next generation MeV gamma-ray detector \citep[e.g.,][]{2016SPIE.9905E..6EW, 2017ICRC...35..798M, 2018JHEAp..19....1D} would be sensitive enough to measure this emission and serve as a critical test to our model.  

\section{Discussion and Conclusions}\label{DC}
\begin{figure}[htbp]
\centering
\includegraphics[width=1\columnwidth]{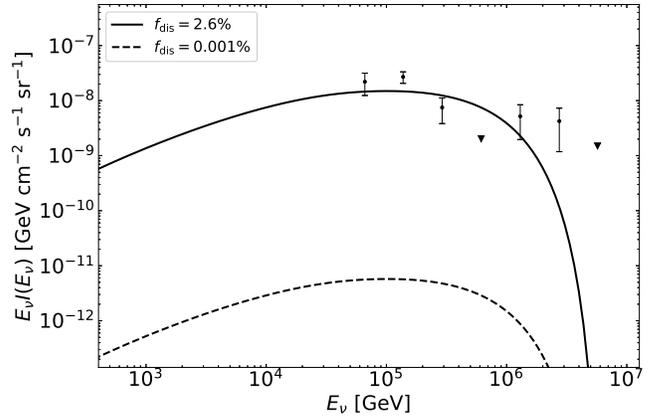}
\caption{Contribution of the blazar ``orphan'' neutrino flare to the all-sky diffuse neutrino background. Green circles represent the measured intensity  of the high-energy $\nu_\mu+\bar{\nu}_\mu$ taken from \cite{2017ICRC...35.1005H} and green triangles indicate the upper limits. The solid curve represents the case with $f_{\rm dis}=2.6\%$ following the contrast between the gamma-ray flux and neutrino flux of TXS~0506+056. The dashed curve shows the resulting diffuse neutrino flux with $f_{\rm dis}=0.001\%$. See Section~\ref{DC} for detailed discussion.
\label{fig:diff}}
\end{figure}

In Section~\ref{results}, we have shown that the inner--blob model may in principle explain the 14-15 neutrino flare from TXS~0506+056 and IceCube-200107A from 4FGL~J0955.1+3551, provided that the inner blob occurs in the X-ray corona around the SMBH. One may then raise a question: how much could such a phenomenon contribute to the all-sky diffuse neutrino background? The answer highly depends on how frequently such a phenomenon could take place in a blazar, and is actually unknown without further understanding on its formation mechanism. Let us parameterize the occurrence probability of certain dissipative process in the X-ray corona of a blazar's SMBH to be $f_{\rm dis}$,  
the diffuse neutrino flux from the entire population of BL Lacs can be estimated as \citep{2014ApJ...793L..18T}
\begin{equation}
E_{\nu}I(E_{\nu}) =\frac{f_{\rm dis} cE_{\nu}^2}{4\pi H_0}\int_{0}^{z_{\rm max}}\int_{L_{\gamma, 0}}^{L_{\gamma, 1}}\frac{j[L_{\gamma}, E_{\nu}(1+z), z]}{\sqrt{\Omega_{\rm m}(1+z)^3+\Omega_{\rm{\Lambda}}}}dL_{\gamma}dz,
\end{equation} 
where $E_{\nu}$ is the observed neutrino energy, $L_{\gamma}$ is the observed $\gamma$-ray luminosity in the range $10^{44}<L_{\gamma}/\rm erg~s^{-1}<10^{46}$ and $j[L_{\gamma}, E(1+z), z]=\Sigma (L_{\gamma}, z)\frac{L_{\nu}(E_{\nu})}{E_{\nu}}$ is the luminosity-dependent comoving volume neutrino emissivity. $\Sigma (L_{\gamma}, z)$ is derived using the luminosity function and parameters for its luminosity-dependent evolution for BL Lacs provided by \cite{2014ApJ...780...73A}. We assume the neutrino luminosity $L_{\nu}$ and the gamma-ray luminosity $L_{\gamma}$ scale as $L_{\nu}=\xi L_{\gamma}$, where $L_{\nu}$ is the differential luminosity of muon and anti-muon neutrino predicted by our model for the 14-15 neutrino flare of TXS~0506+056 \footnote{Here we do not refer to the case of IceCube-200107A because the true neutrino flux/luminosity is of large uncertainty given only one event being detected.} as shown in the upper panel of Fig.~\ref{fig:sed}. Comparing the measured gamma-ray flux and the predicted neutrino flux we have $\xi=2.5$. We then find that $f_{\rm dis}=2.6\%$ is needed in order to explain the diffuse neutrino flux measured by IceCube \citep{2017ICRC...35.1005H} as shown in Fig.~\ref{fig:diff}. It would immediately rule out the possibility of such kinds of blazar ``orphan'' neutrino flares as the main source of diffuse neutrino flux, because the inferred $f_{\rm dis}$ would lead to a very low source density of $<10^{-9}\rm Mpc^{-3}$ and hence violates the constraint from non-detection of neutrino multiplets in the diffuse neutrino events \citep{2014PhRvD..90d3005A, 2016PhRvD..94j3006M}. It suggests that this kind of blazar ``orphan'' neutrino flare must be very rare, implying that dissipations at the jet base cannot occur often or sustain for a very long time. If a dissipative process occurs at an equal probability per unit length along the jet, the probability of forming a blob within the X-ray coronal ($r\lesssim 10^{-4}\,$pc) would be $f_{\rm dis}\sim 0.001\%$ for a  (the inner, highly relativistic) jet length of $\sim$10\,pc, resulting in a subdominant contribution to the diffuse neutrino background (see Fig.~\ref{fig:diff}). Nevertheless, the X-ray corona of SMBH could be a promising neutrino production sites provided that protons are accelerated in the region. For example, \citet{2019ApJ...880...40I} and \citet{2020PhRvL.125a1101M} suggest that protons could be accelerated via plasma turbulence or magnetic reconnection in the corona of AGN and provide a steady neutrino emission. Different from our model, the neutrino emission proposed by these authors is not beamed so the neutrino flux of each source is low but on the other hand the cumulative neutrino flux is compensated by the large population of the sources and persistence of the emission, which may explain the diffuse neutrino flux above 10\,TeV. 

To summarize, we studied the multi-messenger emissions of blazar TXS~0506+056 in the period of the 2014-2015 neutrino flare and blazar 4FGL~J0955.1+3551 during the conjectural emission period of the neutrino event IceCube-200107A  with the inner--outer blob model. Considering that certain dissipative process occurs at the jet base, an inner blob may form within the hot corona around the SMBH and produce a neutrino flare via the $p\gamma$ interactions with the X-ray radiation of the corona, while the non-flare-state multi-wavelength flux of the blazar measured in the same period is ascribed to an outer blob which is far from the SMBH. We found that, on the premise that the multi-wavelength SED being {generally} reproduced, the 14-15 neutrino flare of TXS~0506+056 could be interpreted in a wide range of parameter space, showing the feasibility that the multi-messenger emission of TXS~0506+056 in the period of the 14-15 neutrino flare can be interpreted in the same physical picture of that for IceCube-170922A. The same model could be also applied to the neutrino event IceCube-200107A and explain its association of another blazar 4FGL~J0955.1+3551. By searching the parameter space, we found that the model could well reproduce the neutrino emission of these blazars with an amount of parameter sets. A robust feature predicted by our inner--outer blob model is a simultaneous MeV gamma-ray flare generated by the EM cascade emission with the comparable flux level to that of neutrinos, in agreement with some other models for the 14-15 neutrino flare of TXS~0506+056. In the future, sensitive MeV gamma-ray instrument may be able to catch the MeV flare around the arrival time of a neutrino event from a blazar and serve as a critical test to the inner--outer blob model for the blazar's ``orphan'' neutrino flares with simultaneous multi-wavelength observation.

\section*{Acknowledgements}
We are grateful to the anonymous referee for the invaluable comments and suggest which significantly improves the quality of this work. This work is supported by NSFC grants 11625312 and 11851304, and the National Key R \& D program of China under the grant 2018YFA0404203.

\end{document}